\documentclass[aps,prl,superscriptaddress,twocolumn]{revtex4-1}

\usepackage{bm} \usepackage{graphicx} 
\usepackage{amsmath}
\usepackage{amsthm,stmaryrd,mathtools}
\usepackage{amssymb} 
\usepackage{epstopdf} 
\usepackage{txfonts}
\usepackage{graphicx}
\usepackage{siunitx}
\usepackage{hyperref}
\usepackage{braket}
\usepackage{amsmath}

\usepackage{graphicx,graphics,times,color}
\usepackage{bm}
\usepackage{longtable}
\usepackage{color}
\usepackage{lipsum}
\usepackage{graphicx}
\usepackage{epstopdf}

\def\be{\begin{equation}}
\def\ee{\end{equation}}
\def\ba{\begin{eqnarray}}
\def\ea{\end{eqnarray}}
\def\la{\langle}
\def\ra{\rangle}

\begin{document}

\title{A quantum phase transition detected through one dimensional ballistic conductance}

\author{Abolfazl Bayat}
\affiliation{Department of Physics and Astronomy, University College London, London WC1E 6BT, United Kingdom}

\author{Sanjeev Kumar}
\affiliation{Department of Electronic and Electrical Engineering, University College London, London WC1E 7JE, United Kingdom}
\affiliation{London Centre for Nanotechnology, 17-19 Gordon Street, London, WC1H 0AH, United Kingdom}

\author{Michael Pepper}
\affiliation{Department of Electronic and Electrical Engineering, University College London, London WC1E 7JE, United Kingdom}
\affiliation{London Centre for Nanotechnology, 17-19 Gordon Street, London, WC1H 0AH, United Kingdom}

\author{Sougato Bose}
\affiliation{Department of Physics and Astronomy, University College London, London WC1E 6BT, United Kingdom}

\date{\today}

\begin{abstract}
A quantum phase transition is an unequivocal signature of strongly correlated many-body physics. Signatures of such phenomena are yet to be observed in ballistic transport through quantum wires. Recent developments in quantum wires have made it possible to enhance the interaction between the electrons. Here we show that hitherto unexplained anticrossing between conduction energy sub-bands, observed in such experiments, can be explained through a simple yet effective discretised model which undergoes a second order quantum phase transition within the Ising universality class. 
Accordingly, we observe how the charge distribution, transverse to the direction of the wire will vary across the phase transition.
We show that data coming from three different samples with differing electron densities and gate voltages show a remarkable universal scaling behavior, determined by the relevant critical exponent, which is only possible near a quantum phase transition. 
\end{abstract}


\maketitle

\emph{Introduction.--}  A class of phenomena epitomizing many-body strongly correlated physics is Quantum Phase Transitions (QPT)  and their corresponding universal scaling \cite{sachdev2007quantum,dutta2015quantum}. The observations of QPTs have a long history in bulk magnetic materials \cite{schroder2000onset} and, in recent decades, with ultra cold atoms in optical lattices \cite{greiner2002quantum}. It is fascinating to find evidences of a QPT in an entirely new class of systems as it opens up a new arena for studying strongly correlated physics.  Such strongly interacting many-body systems can also be exploited for practical applications \cite{antonio2015self}. Perhaps the most notable of QPTs has already been observed in semiconductor bulk materials \cite{mott1968metal,imada1998metal}, but there is not yet any observation of such phenomenon in one-dimensional (1D) transport. 

One dimensional strongly correlated systems such as Luttinger liquid \cite{imambekov2009universal,khodas2007fermi,pereira2009spectral}
and Wigner crystal \cite{meyer2008wigner,meyer2007transition,schulz1993wigner,averin1993tunneling,glazman1992quantum,matveev2004conductance} have been theoretically investigated in tunneling conductance measurements. So far, 1D ballistic conductance results \cite{wharam1988one,van1988quantized}, in the regime of being integer multiples of $2e^2/h$, seem to be largely explained by noninteracting electrons. In fact, interactions have only been considered for explaining fractional conductance plateaus such as the $0.25$ \cite{chen2008bias} and the $0.7$  \cite{thomas1996possible,cronenwett2002low,dicarlo2006shot,burke2012extreme,bauer2013microscopic,matveev2004conductance,berggren2002effects} structures. A new anomalous behaviour has been discovered in quasi-1D quantum wires in which the electron-electron interactions can be tuned up  \cite{kumar2014many,hew2009incipient,smith2009row}. As the interaction increases, the first  plateau at $2e^2/h$ initially weakens and then by further increasing the interaction it revives. This manifests itself in an anticrossing between the two lowest energy sub-bands \cite{kumar2014many}. However, there is no satisfactory explanation for this behaviour to date. One cannot appeal to back scattering as the conductance plateaus always remain integer multiples of $2e^2/h$. Reduction to an effective one-body problem through density functional theory \cite{owen2016ground},  only accounts for the narrowing of the first plateau but does not explain its revival. Spin physics in a two-body scenario \cite{huang2015quantum} has also been attempted, but it does not explain why the same anomaly is still present in spin polarized circumstances \cite{kumar2014many}.  In fact, integer plateaus strongly indicates that the Landauer-B\"{u}ttiker \cite{datta1997electronic} paradigm should still be applicable with the sub-bands modified to be strongly correlated many-body states.  It may be noted that, the transition from 1D to quasi-1D has been theoretically predicted in quantum wires, for both classical \cite{meyer2008wigner} and quantum \cite{meyer2007transition,mehta2013zigzag} regimes, the latter including an Ising QPT, but no signature of the ``critical point'' through conductance has been either pointed out or observed.  

\begin{figure*} \centering
	\includegraphics[width=13cm,height=4.5cm,angle=0]{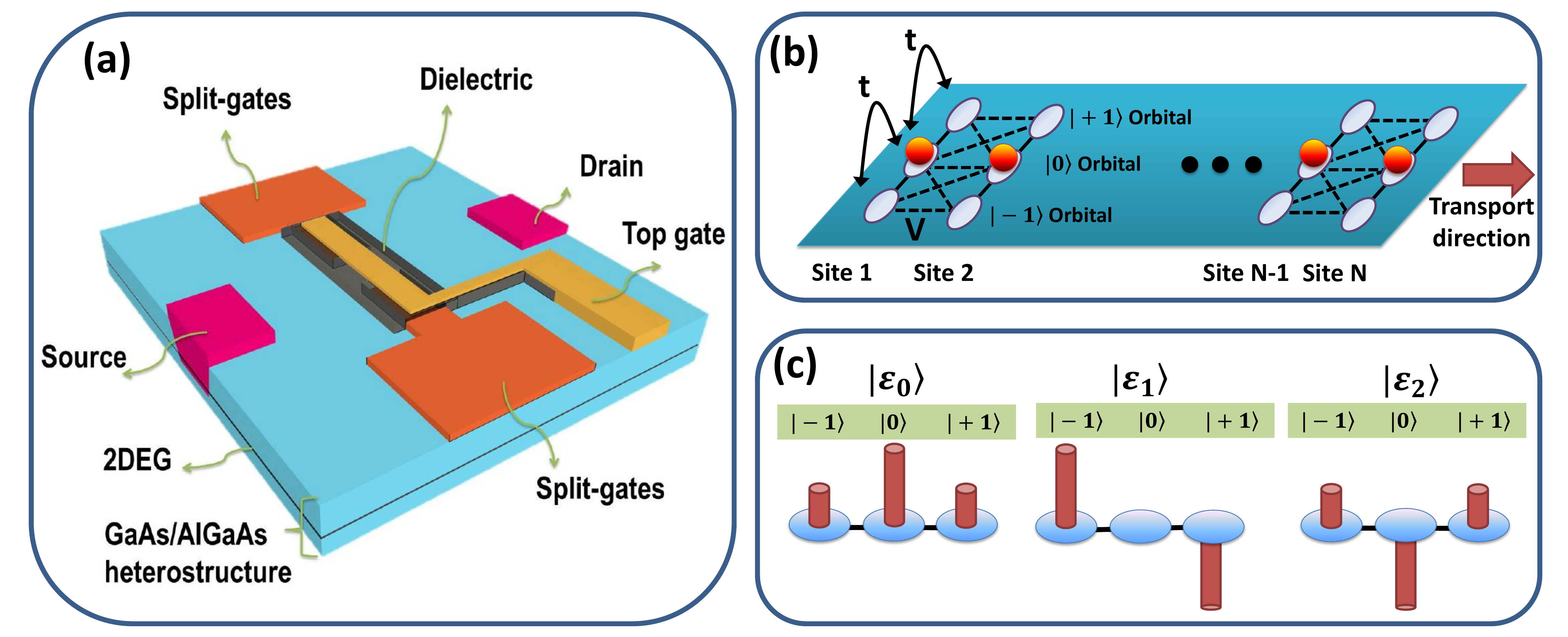}
	\caption{ \textbf{Schematics of the experimental device and theoretical model. (a)} The GaAs$/$AlGaAs heterostructure device consists of a pair of split-gates to define a constriction for electrons from source to drain and a top gate to control the electron density. The split gates are $400$ nm long and $700$ nm wide, and the top gate covers the entire split-gates coving an area of 1 $\mu \text{m}$. The mobility in the dark (light) is estimated to be $1.2  \times 10^6$ $\text{cm}^2/Vs$ ($3.5\times10^6$ $\text{cm}^2/Vs$) and the electron density is expected to be $9\times10^{10}$ $\text{cm}^{-2}$ ($2\times10^{11}$ $\text{cm}^{-2}$). \textbf{(b)} The three orbital model in which the electrons can only tunnel between the transverse orbitals and interact capacitively with neighboring sites as shown by dotted lines.  \textbf{(c)} The three eigenstates of the Hamiltonian (\ref{Hssite}) resembling the spectrum of a harmonic oscillator. The cylinders represent the coefficients of the orbitals in the wave function. }
	\label{fig1}
\end{figure*}

In this letter, we develop a discretized model to explain the anomalous behavior in conductance measurements in a quasi-1D quantum wire. Our model suggests that the observed anomaly is a finite size signature of a QPT which can be characterized via scaling behaviour of conductance data.

\emph{Experiments.--} We have used three different devices, let's call them $S_1$, $S_2$ and $S_3$, with the same geometrical design. A schematic picture of the devices is shown in Fig.~\ref{fig1}(a). 
By applying negative voltage on the split-gates a constriction is created for the electrons flowing from the source to the drain. The electrons pass ballistically through the constriction and give rise to a quantised conductance with each 1D sub-band contributing a conductance of $2e^2/h$ and total conductance is  $2ke^2/h$, where $k{=}1,2,3,...$. The top gate is used to vary the carrier concentration in the quantum wire. 
In all the samples, the 2D electron density ranges from $n_{2D}=9.0\times 10^{10}-2.1\times 10^{11}$ $cm^{-2}$. Using these densities for our 1D wire of length $400$ nm, we can estimate that the number of electrons ranges within  $N{\simeq} 12{-}19$.

\begin{figure}[htb]
	\centering
	\begin{tabular}{@{}cc@{}}
		\includegraphics[width=.24\textwidth]{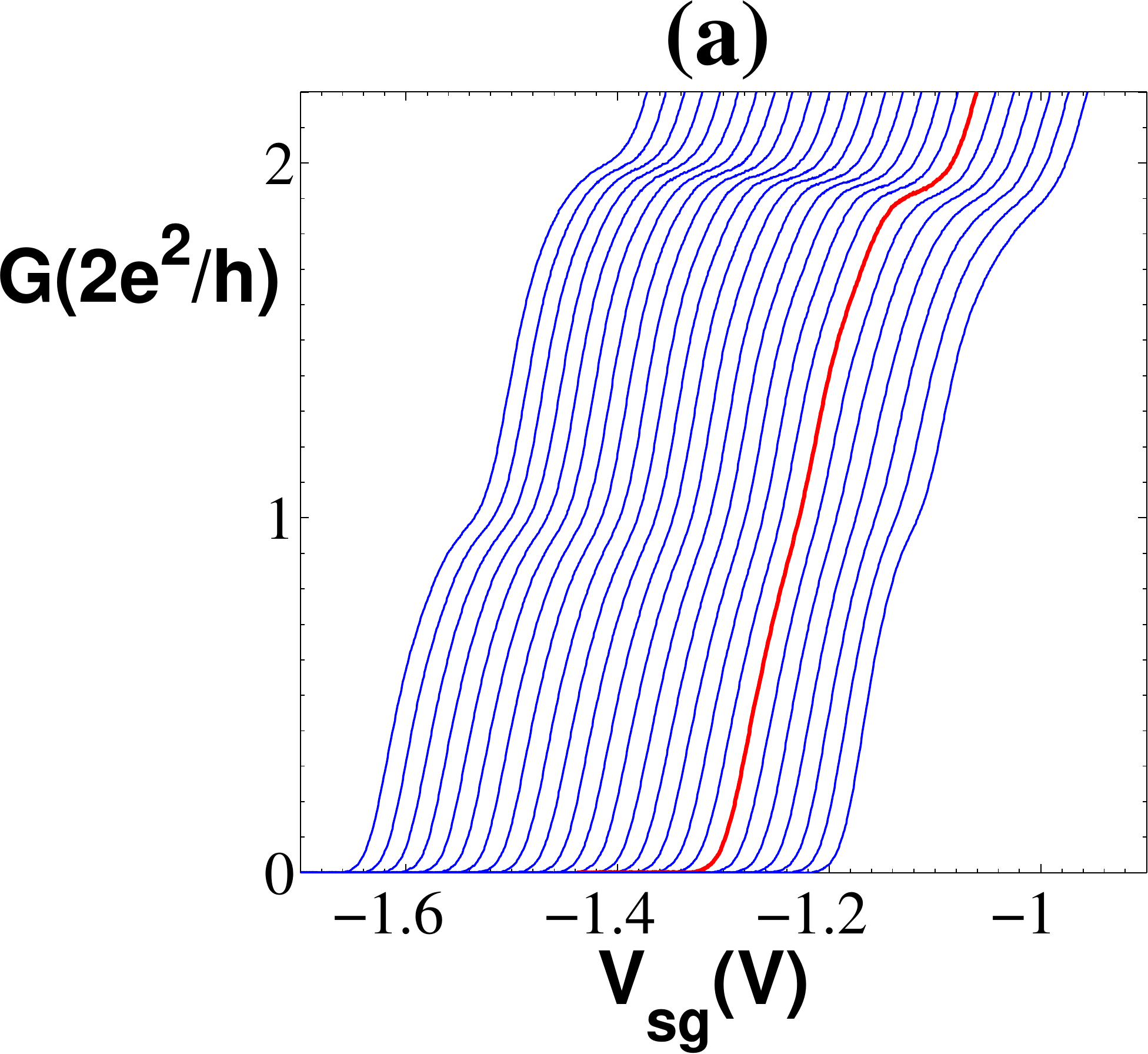} &
		\includegraphics[width=.55\textwidth]{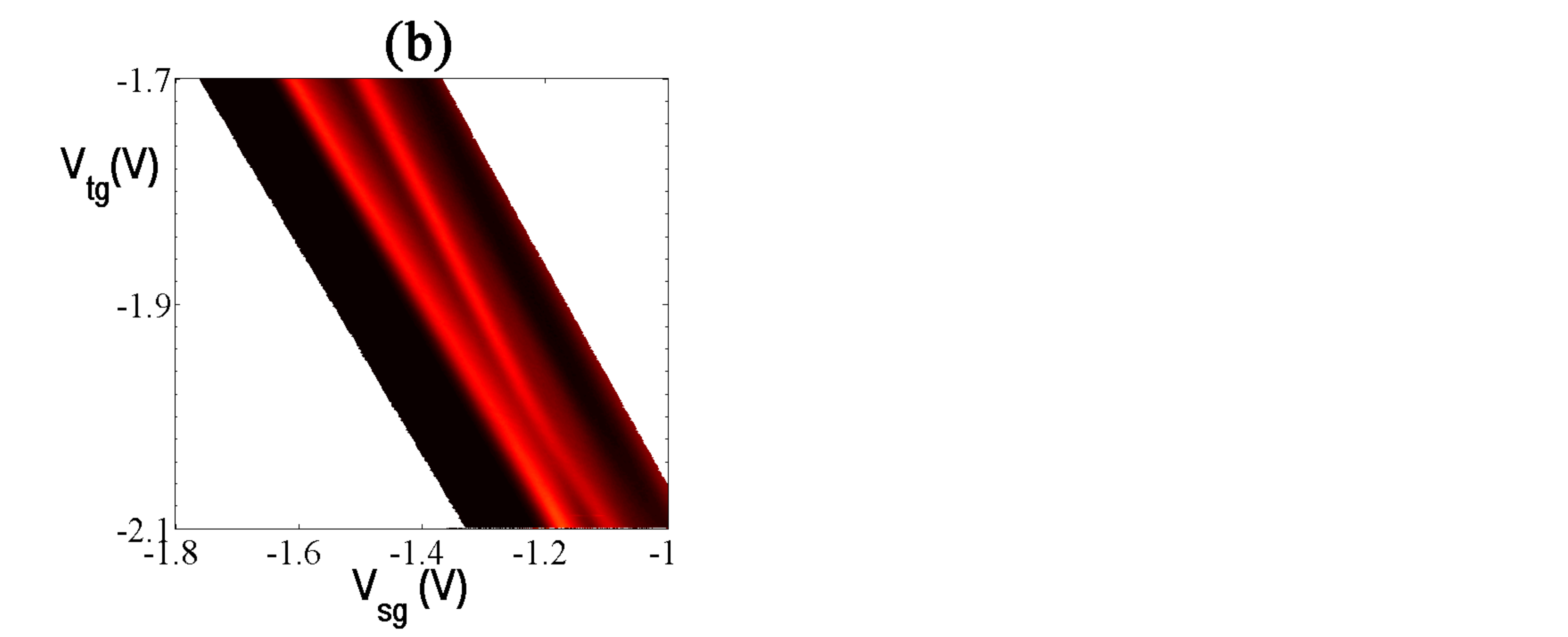} \\		
	\end{tabular}
	\caption{\textbf{Conductance measurements (Experiments). (a)} The conductance, in the unites of $2e^2/h$, measured for the sample device $S_1$ in terms of $V_{sg}$ for various values of $V_{tg}=-1.7$ V (left) to $V_{tg}=-2.1$ V (right). By decreasing $V_{tg}$ the first plateau is weakened until some intermediate values of $V_{tg}^c\backsimeq -2.05$ (red curve) and then is restored by further deacreasing $V_{tg}$. \textbf{(b)} Colour plot of transconductance ($dG/dV_{sg}$) as a function of $V_{sg}$ and $V_{tg}$. The two curves represent the two lowest channel energies.   }
		\label{fig2}
 \end{figure}

We performed  two-terminal differential conductance $G$ measurements using an ac excitation voltage of $10$ $\mu\text{V}$ at $73$ Hz in a cryofree dilution refrigerator with an electron temperature of $70$ mK. Fig.~\ref{fig2}(a) shows the conductance characteristics of $S_1$ as a function of split-gate voltage $V_{sg}$ for various top-gate voltages $V_{tg}$ from $-1.7$ V (left) to $-2.1$ V (right). It is clear from the figure on the left side ($V_{tg}=-1.7$ V) the usual conductance plateaus at integer values of $2e^2/h$ are observed. By lowering the electron density through decreasing $V_{tg}$ and relaxing the confinement via increasing $V_{sg}$ the first conductance plateau $2e^2/h$ is weakened as we move towards the right side of Fig.~\ref{fig2}(a) till we reach the red curve at $V_{tg}^c\backsimeq -2.05$ V. By further decreasing the electron density, i.e. making $V_{tg}$ more negative, the  $2e^2/h$ plateau reappears and gets strengthened. Fig.~\ref{fig2}(b) is the colour plot of transconductance ($dG/dV_{sg}$) as a function of $V_{sg}$ and $V_{tg}$ for the data in Fig.~\ref{fig2}(a), showing what appears as an anticrossing of the ground state and the first excited state. The two states anticross at $V_{tg} = -2.05$ V. This remarkable observation of the weakening and restrengthening of the first conductance plateau is also observed in $S_2$ and $S_3$ but with different values of $V_{sg}$ and $V_{tg}$. The anticrossing feature of Fig.~\ref{fig2}(b), as we show in the following, is linked to a second order quantum phase transition as the interaction between the electrons in the wire is increased by tuning $V_{tg}$ and $V_{sg}$. 


\emph{Theoretical model.--} Since the typical distance ($\sim 50$ nm) between the electrons in the quantum wire  is much longer than the Bohr radius ($\sim 10$ nm) of electrons in GaAs they form an incipient Wigner lattice \cite{meyer2008wigner}.  For simplifying the interaction we use a discretised version of a harmonic oscillator \cite{yung2005perfect,christandl2004perfect} in the transverse direction which needs a minimum of three sites. Thus, for the sake of simulations, we restrict the electrons to reside in one of the three transverse orbitals, i.e. one central orbital $|0\ra$ and two near the edges $|\pm1 \ra$, at each site as shown in Fig.~\ref{fig1}(b). While the electrons in the wire are free to hop in the transverse direction with tunneling $t$, their interaction in the longitudinal direction is only via Coulomb repulsion with no longitudinal tunneling being allowed. 
Similar to Landauer-B\"{u}ttiker theorem \cite{datta1997electronic}, the longitudinal wave function of electrons is a plane wave $e^{-ikx}$ which is factorized out and thus is not explicitly included in the following analysis. The other plane wave with $-k$ is not conducting due to a small $ac$-bias which is used for measuring conductance.  This picture of electron movement through the wire is the same qualitative picture as presented in Ref.~\cite{meyer2008wigner}. Note that the interchannel scattering between the two conducting channels do not have any effect here because of strict conservation laws in one-dimension \cite{yang1967some}.
The single site Hamiltonian for the electron at site $j$ is
\begin{equation}\label{Hssite}
   H_j^{site}=-t \Big( |0_j\ra \la +1_j| + |0_j\ra \la -1_j| + h.c. \Big)+3t/2
\end{equation}
where $t$ is the tunneling between the transverse orbitals and $|\sigma_j\ra$ (with $\sigma{=}0,\pm1$) represents the orbital of the electron at site $j$. The last term is simply an energy shift so that the three eigenvalues of $H_j^{site}$ take the form of a harmonic oscillator as $\varepsilon_n=(n+1/2)t$ (for $n=0,1,2$). The corresponding eigenstates are
\begin{eqnarray}\label{E_n_ vectors}
	|\varepsilon_0\ra&=&\left(|-1\ra+\sqrt{2}|0\ra+|+1\ra\right)/2 \cr
	|\varepsilon_1\ra&=&\left(|-1\ra-|+1\ra\right)/\sqrt{2} \cr
	|\varepsilon_2\ra&=&\left(|-1\ra-\sqrt{2}|0\ra+|+1\ra\right)/2.
\end{eqnarray}
As one can see the charge configurations of the discrete model are independent of $t$ and resemble the wave functions of harmonic oscillators as depicted in Fig.~\ref{fig1}(c). The Coulomb interaction between the electrons can be considered as
\begin{equation}\label{Hint}
   H_{j,j+1}^{int}= \sum_{\sigma,\tau=0,\pm1} \frac{V}{\sqrt{1+|\sigma-\tau|^2}}|\sigma_j,\tau_{j+1}\ra \la \sigma_j,\tau_{j+1}|
\end{equation}
where $V$ is the strength of the Coulomb interaction and the denominator accounts for the distance between the orbitals in adjacent sites, assuming equal horizontal and vertical distances between the orbitals. The total Hamiltonian becomes 
\begin{equation}\label{Hssite}
   H=\sum_{j=1}^{N-1} H_{j,j+1}^{int} + \sum_{j=1}^N H_j^{site}
\end{equation}
where $N$ is the number of electrons in the wire and is taken to be fixed as by exiting one electron from the wire another one gets in.  
In the noninteracting regime, i.e. $V=0$, the three lowest transverse conducting channels  (or sub-bands) are
\begin{equation}\label{En_V0}
|\Psi_n(V\!=\!0,t)\ra=|\varepsilon_n\ra^{\otimes N}, \quad  E^{ch}_n(V\!=\!0,t)=(n+1/2)Nt,
\end{equation}
where $n\!=\!0,1,2$. It is worth mentioning that the Hamiltonian $H$ in Eq.~(\ref{Hssite}) has $3^N$ eigenvalues  but according to  Landauer-B\"{u}ttiker model for ballistic transport only three of them, given in Eq.~(\ref{En_V0}), are relevant for conductance measurements. In order to capture more conducting channels one has to increase the number of transverse orbitals. 

\begin{figure} \centering
    \includegraphics[width=.5\textwidth]{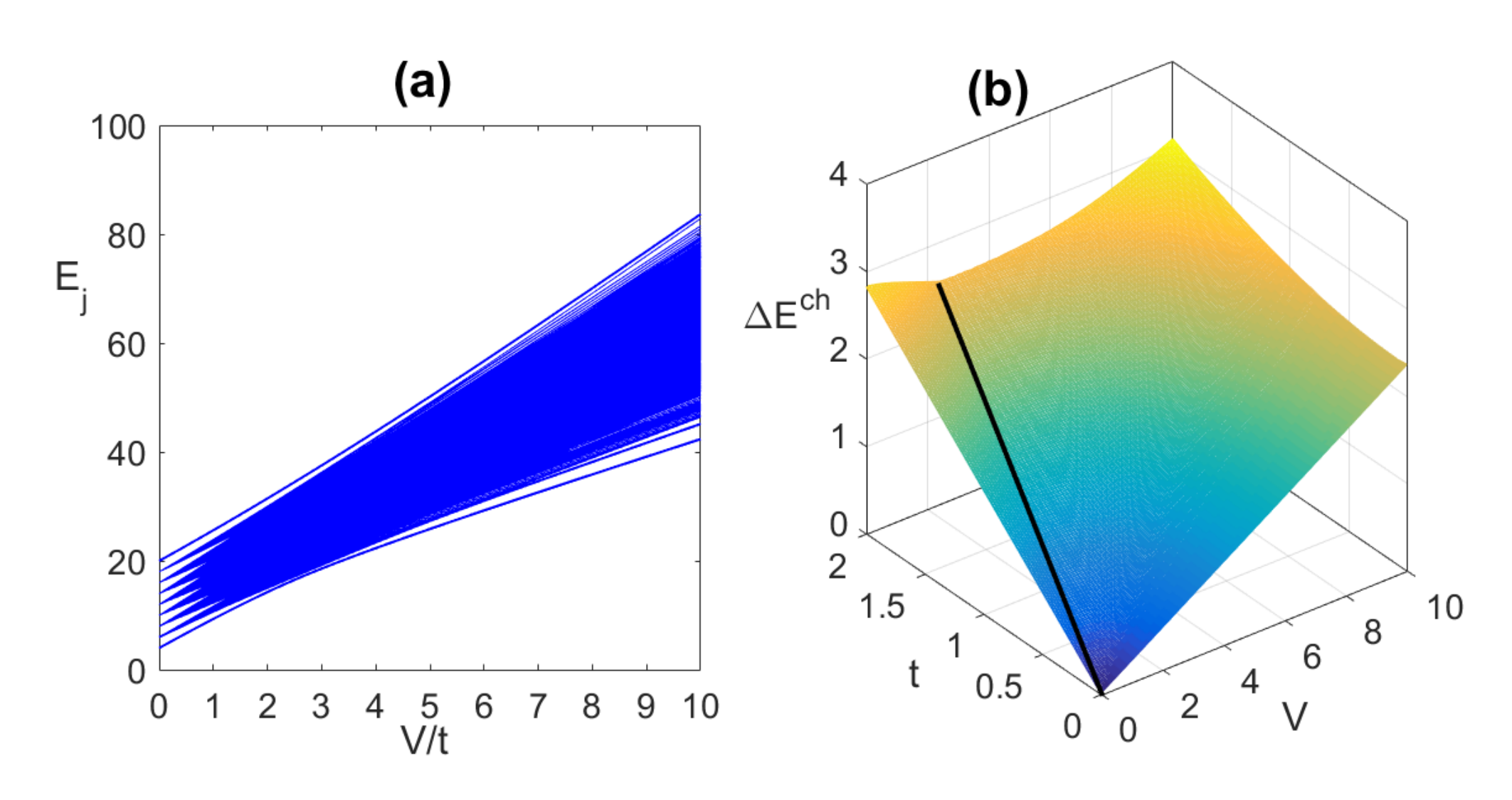}
    \caption{ \textbf{Spectrum analysis (Theory).}  \textbf{(a)} The energy spectrum as a function of $V/t$ for a chain of length $N=8$. \textbf{(b)} The energy gap between the conducting channels versus confinement energy $t$ and Coulomb interaction $V$. The black line shows a local dip in the energy gap.
      }
     \label{fig3}
\end{figure}

To see the effect of interactions, we use adiabatic continuation by gradually turning on the interaction and evolve the non-interacting levels of Eq.~(\ref{En_V0}). The whole energy spectrum of the Hamiltonian $H$ is plotted in Fig.~\ref{fig3}(a) as a function of $V/t$. As we can see at the weak interaction regime there is a clear band structure in the system making the whole system gapped. By increasing $V/t$ the whole spectrum shrinks into a single band around $V/t\simeq 1.6$ which is expected to approach 1 as $N$ increases. Increasing the interaction even further opens the gap in the system again suggesting the presence of a second order quantum phase transition between two gapped phases through a critical gapless point. We have to be careful that the conducting channels are not the two lowest eigenstates of the system. In Fig.~\ref{fig3}(b), we depict  $\Delta E^{ch}{=}E^{ch}_1{-}\!E^{ch}_0$ as a function of $t$ and $V$ for a system of length $N{=}8$. The remarkable feature is that there is always a local minimum in $\Delta E^{ch}$ for any path going from non-interacting to strong interaction limit in the $t{-}V$ plane, depicted as a black line in Fig.~\ref{fig3}(b).

\begin{figure} \centering
    \includegraphics[width=.45\textwidth]{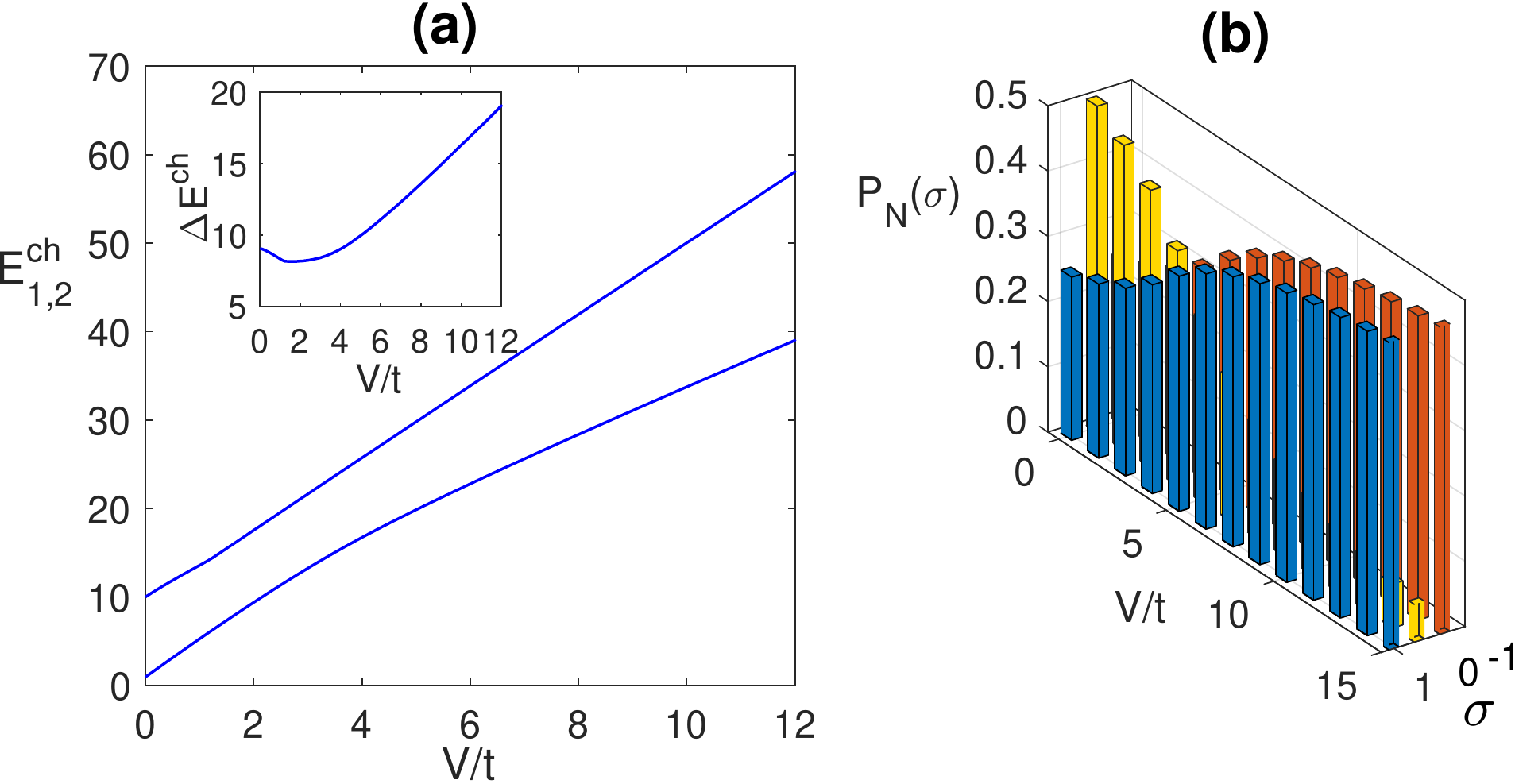}
    \caption{ \textbf{Energy anticrossing and electronic wave functions (Theory). (a)}  The two lowest energy levels of the two conducting channels versus $V/t$ in a system of length $N=8$. The inset shows the energy gap versus $V/t$. \textbf{(b)} The electron density in each orbital at the exit site $N$ as a function of interaction $V/t$.  }
     \label{fig4}
\end{figure}

To have a better insight, in Fig.~\ref{fig4}(a) we plot the $E_1^{ch}$ and $E_2^{ch}$ as functions of $V/t$ for a system of size $N{=}8$. An anticrossing at $V/t\backsimeq 1.6$ is evident which is qualitatively similar to the anticrossing observed in experimental data shown in Fig.~\ref{fig2}(b). To see this even more explicitely we plot $\Delta E^{ch}$ as a function of $V/t$ in the inset of Fig.~\ref{fig4}(a).  

The three-orbital model can also show the charge configuration in each conducting channel, in particular, the ground state $|E^{ch}_0\ra$.  While for $V=0$, the electron wave function in the ground state has a central dominant peak, by increasing $V/t$ the electrons are pushed towards a zig-zag configuration
\begin{equation}\label{zig-zag} \nonumber
  |\Psi_0 (V \rightarrow \infty)\ra=\left( |+1,-1,...,-1\ra+|-1,+1,...,+1\ra \right)/\sqrt{2}.
\end{equation}
To visualize the charge configuration of the output current we compute the probability of finding the electron in each of the three orbitals at site $N$
\begin{equation}\label{Prob_N}
  P_N(\sigma)= \la \Psi_0 | \Big( \mathcal{I} \otimes |\sigma_N\ra \la\sigma_N | \Big) |\Psi_0 \ra
\end{equation}
where $\mathcal{I}$ stands for identity operator acting on sites $1$ to $N-1$ and $\sigma=0,\pm 1$. In Fig.~\ref{fig4}(b) we plot $P_N(\sigma)$ as a function of $V/t$ in a system of size $N{=}8$. As it is clear from the figure, the central peak in the charge configuration which is initially dominant in the non-interacting regime disappears in the extreme interacting regime. In the transition point where $V/t\backsimeq 1$ the three peaks are almost the same and in the regime of extreme high interaction only the edge peaks remain prominent. 

It is worth mentioning that a QPT occurs at the thermodynamic limit in which $N{\rightarrow} \infty$. While by increasing $N$ the anti-crossing becomes sharper, our theoretical model with only $N=8$ electrons is justified as: (i) it is estimated to have ${\sim} 10$ electrons in our quantum wire during the experiments and thus our analysis provides a finite size precursor of a QPT for the devices at hand; (ii) with finite size scaling, which will be discussed in the next section, one can deduce thermodynamic behavior using data of very finite systems; (iii) numerically we need the whole spectrum of the system to extract the information for conducting channels which makes us limited to $N=8$ electrons as the Hilbert space grows exponentially (i.e. $3^N$). Indeed, we believe this is the best that can be done with current computational power. 

\begin{figure} \centering
    \includegraphics[width=.5\textwidth]{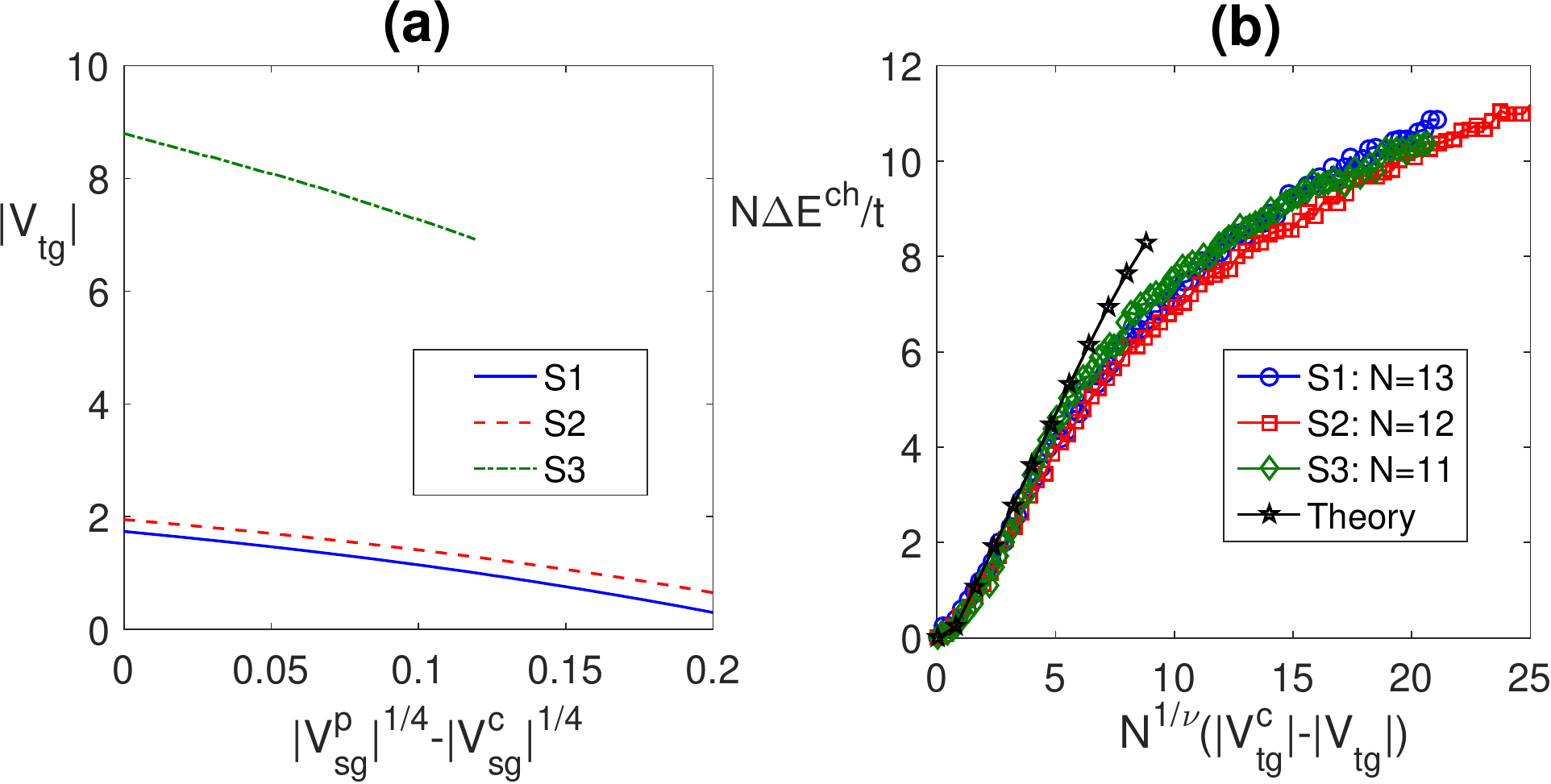}
    \caption{ \textbf{Scaling (Experiments).}  (a) The top gate voltage $V_{tg}$ as a function of $|V_{sg}^p|^{1/4}$, with an irrelevant shift to bring the curves near each other.  \textbf{(b)} The data collapse for the data measured in all the three samples and our theory prediction. }
     \label{fig5}
\end{figure}

\emph{Scaling.--} In this section we try to connect our theoretical model with experimental data. In the experiment the only control parameters are $V_{tg}$ and $V_{sg}$. By applying negative voltages on the top gate the electron density decreases as
\begin{equation}\label{density}
n=n_0-\frac{\epsilon_0\epsilon_r}{ed}|V_{tg}|
\end{equation}
where $n_0$ is electron density in the absence of top gate voltage, $\epsilon_0$ is the vacuum permittivity, $\epsilon_r$ is the GaAs dielectric constant, $e$ is electron charge, $d=500$ nm is the distance between top gates and the 2DEG  and finally $|\cdot|$ stands for absolute value. The kinetic energy of free electrons outside the wire will be $\hbar^2n^2/2m$. For any value of $V_{tg}$ there exists a particular split-gate voltage $V_{sg}^p$ for which the wire is pinched off and the current stops. At the pinch off one can write
\begin{equation}\label{Vsg_pinch}
  eV_{sg}^p=\frac{1}{2}m^*\omega^2A^2
\end{equation}
where $A$ is the width of the potential, $\omega$ is the strength of the confinement which determines the ground state energy $\hbar \omega/2$ and $m^*$ is the electron effective mass. At the pinch off, the whole kinetic energy of the free electrons is transferred to the confinement energy with no longitudinal momentum   
\begin{equation}\label{identity_energy}
\hbar^2n^2/2m^*=\hbar \omega /2.
\end{equation}
In this equality, by replacing electron density $n$ from Eq.~(\ref{density}) and $\omega$ from Eq.~(\ref{Vsg_pinch}) we get
\begin{equation}\label{Vtg_Vsg_pinch}
  |V_{tg}|=\frac{ed}{\epsilon_0\epsilon_r} \left[ n_0 - \left(\frac{2e m^*}{A^2 \hbar^2} \right)^{1/4} |V_{sg}^p|^{1/4}  \right].
\end{equation}
In Eq.~(\ref{Vtg_Vsg_pinch}), while the electron density $n_0$ may vary from one device to another the coefficient of $V_{sg}^p$ is identical for all three samples. In deriving Eq.~(\ref{Vtg_Vsg_pinch}) we have two implicit assumptions: (i) $V_{sg}$ has no effect on electron density (\ref{density}); and (ii) $V_{tg}$ does not affect the potential of the wire. In Ref.~\cite{owen2016ground}, a detailed analysis for the potential inside the wire as a function of $V_{sg}$ and $V_{tg}$ is provided  which qualitatively supports the above assumptions. While the first assumption may not be very precise, the second one seems alright as the top gate covers the whole wire and thus mainly gives an offset. 
To see the accuracy of Eq.~(\ref{Vtg_Vsg_pinch}), we plot $|V_{tg}|$ as a function of $|V_{sg}^p|^{1/4}$ (with an irrelevant shift to bring the curves near each other) in Fig.~\ref{fig5}(a) for all three samples.  While in all samples $V_{tg}$ linearly varies with  $|V_{sg}^p|^{1/4}$ over a long range of voltages, as predicted by Eq.~(\ref{Vtg_Vsg_pinch}), the curves start to bend for large values of $V_{sg}$ suggesting that the first assumption is not very precise at that limit.

The most important feature of a quantum phase transition is scaling \cite{sachdev2007quantum,dutta2015quantum}. This implies that as $N\rightarrow \infty$, the energy separation between the two lowest energy sub-bands decreases as  $\Delta E^{ch} \sim |\lambda-\lambda_c|^\nu$, where $\lambda=V/t$ is the control parameter, $\lambda_c$ is the critical point and $\nu$ is the critical exponent.
In the theoretical model, in the absence of interaction, the energy separation  between the two lowest sub bands is $t=\hbar \omega$. Moreover, the Coulomb interaction can be written as $V=e^2n/4\pi\epsilon_r\epsilon_0$. Thus, by using Eqs.~(\ref{identity_energy}) and (\ref{density})one can determine $\lambda$ as a function experimental parameters 
\begin{equation}\label{lambda}
  \lambda-\lambda_c=\frac{4\pi (\epsilon_0 \epsilon_r\hbar)^2}{mde^3} \left( |V_{tg}|- |V_{tg}^c| \right) .
\end{equation}

Experimentally, the energy separation between the two lowest sub-bands is $\Delta E^{ch}=eW$, where $W$ is the width of the first plateau of the conductance curve. The standard finite size scaling ansatz \cite{sachdev2007quantum} implies that 
\begin{equation}\label{finite_size_scaling}
  N \Delta E^{ch}/t=f(|\lambda-\lambda_c| N^{1/\nu}).
\end{equation}
where $f(\cdot)$ is an arbitrary function. In Eq.~(\ref{finite_size_scaling}) both $N$ and $\nu$ are free parameters to be found for the data collapse. For each sample, $N$ might be different but $\nu$ should be fixed and independent of the sample. In Fig.~\ref{fig5}(b) we use $\Delta E^{ch}/t = W(V_{tg})/W(V_{tg}=0)$ where $W(V_{tg})$ is the width of the first plateau as a function of top-gate voltage. For obtaining the best data collapse for the three samples we numerically find the optimal values $\nu$ and $N$.
Our data collapse reveals that $\nu=1$. Moreover, the optimal values found for $N$ matching well with realistic values of $N=12-19$ estimated for our samples. 
The remarkable data collapse, observed in Fig.~\ref{fig5}(b), is not only in excellent agreement with the theoretical predictions of our orbital model but also is fully consistent with the results of Refs.~\cite{meyer2008wigner,meyer2007transition} in which an Ising type quantum phase transition (with $\nu=1$) has been used to explain the gapped nature of the zig-zag phase after the QPT. Moreover, the displacement of electrons from a 1D straight line  to a zig-zag configuration can be used as on order parameter for characterizing this QPT. A Monte Carlo simulation of such transition has been studied in Ref.~\cite{mehta2013zigzag}. Nonetheless, unlike conductance the observation of electronic wave function is still a major experimental challenge.


\emph{Conclusion.--} We show that the anomalous weakening of the first plateau in conductance measurements, observed in three different samples with differing gate voltages and densities, is an indicator of the emergence of a second order QPT which can be captured through a universal scaling of conductance data. The critical exponents found from our scaling analysis lies in the Ising universality class which is in agreement with field theory analysis. 

\emph{Acknowledgements.}  The authors acknowledge the EPSRC grant $EP/K004077/1$. SB also thanks the European Research Council for the European Unions Seventh Framework Programme $(FP/2007-2013)$ $/$ ERC Grant Agreement No.~$308253$.


%

\end{document}